# Towards label-free distributed fiber hydrogen sensor with stimulated Raman spectroscopy

**F. Y**ANG[1,2,†], **Y. Z**HAO[1,2,†], **Y. Q**I[1,2], **Y. Z. T**AN[1,2], **H. L. H**O[1,2] AND **W. J**IN[1,2,*]

[1] *Department of Electrical Engineering and Photonics Research Center, The Hong Kong Polytechnic University, Hong Kong, China.*
[2] *Photonic Sensors Research Center, The Hong Kong Polytechnic University Shenzhen Research Institute, Shenzhen 518057, China.*
[†] *These authors contributed equally to this work.*
[*] *eewjin@polyu.edu.hk*

**Abstract:** Hydrogen detection is of great importance in chemical and energy industries. Optical fiber hydrogen sensors show flexibility and compactness, and have potential for distributed analysis. However, traditional fiber sensors encounter a challenge for light to interact with hydrogen directly since hydrogen only display weak quadrupole absorption, and metallic palladium and platinum thin-film coatings are typically used as an optically detectable label. Here, based on stimulated Raman spectroscopy in hollow-core photonic crystal fibers, we investigate the label-free optical fiber distributed hydrogen sensors operating in the optical telecommunication band. The approach of distributed Raman measurement represents a new paradigm in fiber sensors, potentially allowing distributed chemical analysis in gas or liquid phase with high sensitivity and selectivity.

## 1. Introduction

Recently, hydrogen attracts much attention as a clean and inexhaustible source of energy. It has some unique properties such as low minimum ignition energy (0.017 mJ), wide explosion range (4-75%) and high flame velocity. However, it leaks out easily due to the smallest molecule size. Detection of hydrogen sensitively and selectively is important and in urgent need for safety and other applications. Several techniques were reported for hydrogen detection, including those based on catalysis [1], thermal conductivity [2] and resistance [3], as well as electrochemical [4], mechanical [5] and optical properties [6]. Optical sensors are most appropriate for applications in harsh and dangerous environments, owing to their immunity to electromagnetic interference, free from electric sparking, and multiplexed and distributed sensing capability [7].

So far, most optical hydrogen sensors used thin films of palladium or chemochromic oxides coated onto the tip or surface of an optical fiber. Hydrogen concentration is measured indirectly via the hydrogen-absorption induced strain, temperature or refractive index change [8]. One of the most important benefits offered by optical fiber sensors is the distributed sensing capability. Several experiments on distributed hydrogen detection with thin film coatings were conducted [9-11]. Sumida et al. reported multi-point hydrogen detection based on evanescent field absorption of a $Pt/WO_3$ coated fiber by use of optical time domain reflectometry (OTDR) [9]. They demonstrated the measurement of 1% hydrogen in three different locations separated by a minimum distance of a few meters. However, the high optical loss (62 dB/m) of the coated region makes it difficult to realize distributed sensing. Wang et al. demonstrated a distributed hydrogen sensor using acoustically induced travelling long-period gratings (LPG) to measure the heat generated by Pt-assisted combustion of $H_2$ and $O_2$ [10]. However, the acoustic damping along the fiber seriously limits the sensing length. Chen et al. reported distributed hydrogen sensing over 2-m-length of optical fiber and achieved a spatial resolution of 1 cm [11]. They coated single-mode fiber (SMF) with Palladium/Copper and measured hydrogen absorption-induced strain with optical frequency domain reflectometry (OFDR). Electric heating of Palladium via the Copper coating was used to improve detection sensitivity but it would have spark hazard for practical applications.

Raman spectroscopy is a label-free technique for material analysis and it provides the fingerprint by which molecules can be identified. Spontaneous Raman scattering is however very weak and watt-level pump power and liquid-nitrogen-cooled detectors are typically needed to detect the weak scattering light [12]. Coherent anti-Stokes Raman scattering (CARS) and stimulated Raman scattering (SRS) are commonly used as coherent techniques to enhance the weak Raman signal. CARS is a four-wave mixing process, in which a pump beam and a Stokes beam interact with the analytes to generate a signal at the anti-Stokes frequency. However, CARS shows non-resonant background and the CARS signal is proportional to the square of the molecular concentration [13]. SRS has no non-resonant background and the signal is directly proportional to the molecular concentration. SRS could greatly benefit from the newly developed hollow-core photonic crystal fiber (HC-PCF) technology. The long interaction length, smaller beam size and hence higher optical intensity as well as nearly 100% overlap between optical field and gas sample make it an excellent platform for gas sensing based on absorption [14] and non-linear optical interaction [15]. Recently, HC-PCF based Raman gas sensors have been reported [12, 16-19]. Compared with free-space system, the sensitivity of the HC-PCF sensors can be two to three orders of magnitude higher. To the best of our knowledge, no label-free distributed optical fiber hydrogen sensor has been reported.

In this work, we demonstrate the first label-free, distributed optical fiber hydrogen sensors based on backward stimulated Raman scattering in a hollow-core optical fiber with precisely manufactured micro-fluidic channels for gas transportation. The distributed Raman gain measurement approach is highly sensitive with a large dynamic range, and is both time- and spatially-resolved. By selecting the proper Raman transition of hydrogen, the system is operated in the telecommunication wavelength band, allowing the use of compact and cost-effective optical components and possibility of commercializing for practical application.

## 2. Method

SRS is a third-order nonlinear process involving a pump photon, a down-shifted Stokes photon and a specific vibrational or rotational mode of molecules. By propagating a pump ($\omega_P$) and a probe ($\omega_S$) beam in a material and matching their frequency difference $\Delta\omega = \omega_P - \omega_S$ to a molecular vibrational or rotational transition frequency $\Omega$ of the material, the intensity of the probe experiences a gain (stimulated Raman gain, SRG) and the intensity of the pump experiences a loss (stimulated Raman loss, SRL). This pump and probe detection scheme is often termed as the seeded SRS. HC-PCFs have proven to be an excellent platform for SRS [20]. A HC-PCF tightly confines the laser beams inside its core area and provides a long interaction distance between laser beams and gas molecules, enabling distributed gas sensing. Here we use NKT Photonics' HC-1550-06 fiber as the sensing fiber and employ the active Raman transition $S_0(0)$ between rotational level J = 0 and J = 2 of hydrogen molecules, as shown in Fig. 1(a). When the frequency difference between the pump and probe matches the rotational Raman shift $\Omega = 354.36$ cm$^{-1}$, the intensity variation of the probe (Stokes) beam may be expressed as $\Delta I_S \propto g I_S I_P \propto I_S I_P \Delta N / \Gamma$ in the steady state [21]. $I_S$ and $I_P$ are respectively the intensities of the probe and pump. $g$ is the Raman gain factor and $\Gamma$ the Raman linewidth $\Delta N$ is the number density difference of hydrogen molecules between J = 0 and J = 2. $\Delta N$ is proportional to hydrogen concentration, which can be recovered by detecting $\Delta I_S$.

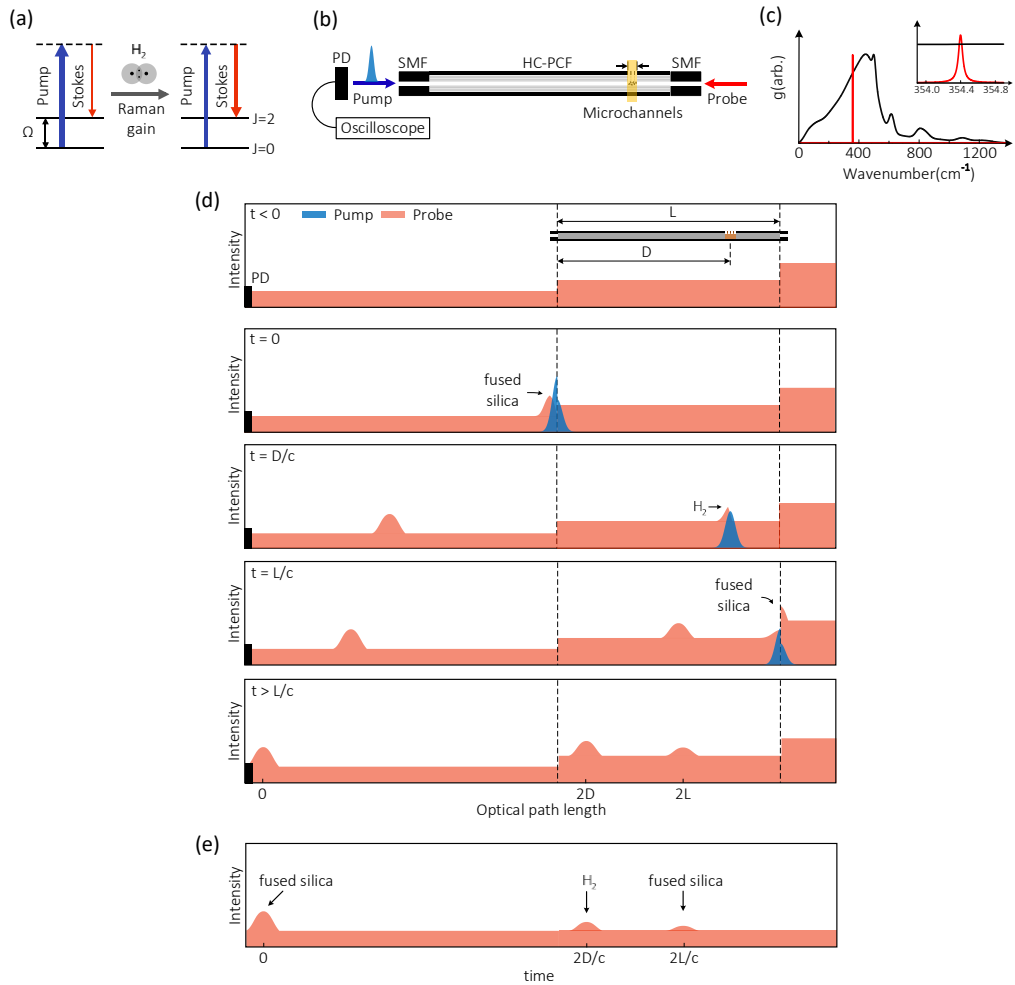

Fig. 1. Distributed hydrogen sensing with backward SRS in a HC-PCF. (a) The energy level diagram of $S_0(0)$ transition of hydrogen. (b) Basic setup for distributed hydrogen detection. A pump pulse and a c.w. probe light enters the HC-PCF sample in the opposite direction. The region with yellow-shade has multiple micro-channels and are for gas ingress/egress. The construction of the HC-PCF sample is described in the Appendix D. The intensity of the probe light is measured by a photodetector (PD) and recorded by an oscilloscope. (c) The backward Raman gain spectrum of $S_0(0)$ transition of hydrogen and fused silica [23]. The inset shows an enlarged view around the $S_0(0)$ transition of hydrogen. (d) Schema showing the principle of distributed hydrogen sensing. The panels (from top to bottom) show the probe and pump intensity distribution (not to scale) along the fibre at different time moment ($t < 0$, $t = 0$, $t = D/c$, $t = L/c$ and $t > L/c$). The step changes of the probe light intensity at the SMF/HC-PCF joints are due to the coupling losses. The yellow-shaded region at the distance $D$ from the left SMF-HC-PCF joint is filled with hydrogen. The gain of probe light is enlarged for clarity purpose. Inset in the top panel shows the HC-PCF sample with micro-channels. Transmission loss of HC-PCF is ignored in the schema. (e) SRG signal detected by photodetector (PD) in the time domain.

In the HC-PCF, the pump could travel in the same direction as the probe (forward SRS), or in the opposite direction (backward SRS). The Raman linewidth $\Gamma$ for the backward SRS is larger due to Doppler broadening [22], resulting in a smaller backward Raman gain factor. The backward and forward Raman gain factors for $S_0(0)$ transition of pure hydrogen at 1 atm and 296 K are calculated to be 4.5 cm/TW and 0.1 cm/GW respectively (Appendix B). However, the forward SRS could not provide the location information needed for distributed sensing. Here, we demonstrate an all-fiber distributed hydrogen sensor based on backward SRS using a pump and probe scheme shown in Fig. 1(b). The narrow Raman gain spectrum of $S_0(0)$ transition of hydrogen lies within the wider Raman gain spectrum of fused silica, as shown in Fig. 1(c). The Raman gain of the silica SMF pigtails could act as markers to identify the beginning and ending of the HC-PCF sensing region.

Raman transition $S_0(0)$ is communication band compatible with the pump located in C band and the Stokes located in L band. We here use a pump around 1532 nm and a probe around 1620 nm, both wavelengths are within the low loss transmission window of the HC-1550-06 fiber (with loss < 30 dB/km from 1490 to 1680 nm). As illustrated in Fig. 1(d), as the pump pulse (blue with pulse width $\tau$) travels along the HC-PCF, it encounters the counter-propagating c.w. probe (red) beam at different locations along the HC-PCF. The group index of the fundamental mode in the HC-PCF is ~1 and hence the probe and the pump propagate approximately at the vacuum speed $c$. The measurement begins when the pump pulse enters the HC-PCF from the SMF pigtail ($t = 0$). The probe beam experiences firstly the SRG of fused silica. Then, the pump pulse travels in the HC-PCF and the probe experiences a SRG of hydrogen if there are hydrogen molecules in the HC-PCF ($t = D/c$). Finally, the pump pulse reaches the output SMF pigtail and the probe experiences the SRG of the output SMF ($t = L/c$). The total optical path length over which the probe beam interacts with the pump pulse is $2L$, corresponding to a duration of $T = 2L/c$ in the time domain. Figure 1(e) shows the variation of the detected probe intensity as a function of time, which maps the concentration of hydrogen along the HC-PCF. The spatial resolution of the SRG measurement would be limited by the pump pulse width $\tau$ and given by $\Delta = \tau c/2$.

## 3. Result

### 3.1. Test of lower detection limit and response time

The lower detection limit of the distributed hydrogen sensing system (Figs. 1(c) and 7 in Appendix C for more details) was evaluated with a 100-m-long HC-PCF (L = 100 m) spliced to SMF pigtails at both ends. Micro-channels were drilled over a 2.2-m-long region at around D = 88 m, which is the yellow shaded region shown in Fig. 1(b). This region with micro-channels is exposed to 4% of hydrogen balanced with nitrogen at a pressure of 1 atm. The pump pulse has a duration of 18-ns and the peak power delivered to the HC-PCF sample was ~30 W, corresponding to a peak pump intensity of 110 MW/cm$^2$. The power of c.w. probe beam is ~100 µW measured before the photodetector, which is much smaller than that of the pump pulse. Therefore, SRL of the pump pulse is small and need not to consider in our experiment. Besides, the depletion of probe light does not disturb the magnitude of hydrogen measurement since the probe light experiences the same loss. But the attenuation of the pump pulse will result in a weaker SRG at a longer distance. Figure 2(a) shows the measured SRG signal over the 100-m-long HC-PCF sample. The gain peaks at 0 m and 100 m are due to SRG of silica SMF pigtails. It gives the information about the beginning and ending of the HC-PCF. The gain peak around 88 m is due to the SRG of hydrogen, which has an amplitude of 1.4 mV. The standard deviation of the noise from 40 to 80 m is 29 µV, giving a noise equivalent concentration (NEC) of 833 ppm. This detection limit is ~50 times lower than the 4% lower explosive limit of hydrogen.

To test the response time of the distributed hydrogen sensor, the region with micro-channels was placed inside a gas chamber with the size of 20 cm × 30 cm × 1 cm. The gas chamber was initially filled with pure nitrogen. Then 20% hydrogen balanced with nitrogen was filled into the gas chamber. After the SRG signal of hydrogen reached a steady value, the gas chamber was purged with pure nitrogen. Figure 2(b) shows the SRG distribution along the 100-m-long HC-PCF as a function of time. The two Raman gain peaks at the position of 0 and 100 m are the SRG of the SMF pigtails. The SRG signal of hydrogen at ~88 m as a function of time is shown in Fig. 2(c). The response time, defined as the rising time to $1-e^{-1}$ of the steady state after exposing to hydrogen, is 46 s. The recovery time, defined as the falling time to $e^{-1}$ of the steady signal after purging with pure nitrogen, is 54 s.

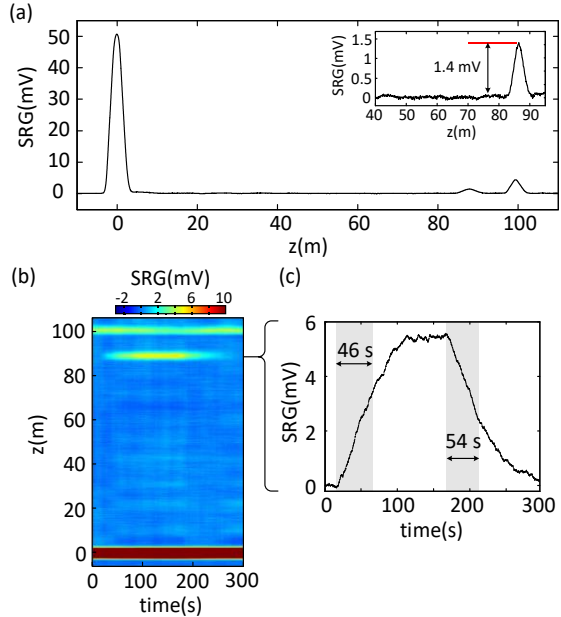

Fig. 2. Stimulated Raman gain trace along a 100-m-long HC-PCF. (a) The measured SRG trace along the 100-m-long sensing HC-PCF with 18-ns pump pulse. The frequency difference between the pump and probe was tuned to the $S_0(0)$ rotational Raman transition of hydrogen(on-resonance). The inset shows the enlarged SRG signal around 88 m. The number of averages used in the oscilloscope is 12,000. (b) The distributed Raman gain trace as a function of time of the hydrogen filling and recovering process. The number of averages is 200. (c) SRG signal at the position of 88 m. The shaded regions show the filling and recovering process, which define the $e^{-1}$ response (rising and recovering) time.

*3.2 Test of dynamic range*

The dynamic range of distributed SRG gas detection system was tested with the same 100-m-long HC-PCF sample. SRG signal was measured when the 2.2-m-long region with micro-channels was filled with different concentrations of hydrogen. The HC-PCF sample was placed inside a gas chamber, which was filled initially with pure nitrogen and then a mixture of pure hydrogen and nitrogen of different ratios (see Appendix A for details about gas preparation). The results are shown in Fig. 3(a) and a good linear relationship is obtained for hydrogen concentration up to 50%. The slope of the linear fit is 22 nV/ppm. The SRG signal of pure hydrogen was also measured and is 17.3 mV, which is ~21% smaller than the value predicted by the linear fit. It shows that our system could be used to detect hydrogen from low concentration to pure hydrogen. The lower detection limit in terms of NEC is 833 ppm, giving a dynamic range of over three orders of magnitude (from 833 ppm to ~100%). The SRG signal as function of peak pump power level is shown in Fig. 3(b). The signal amplitude is linearly proportional to the peak pump power, indicating that further enhancement of detection sensitivity is possible by simply increasing the peak power of the pump.

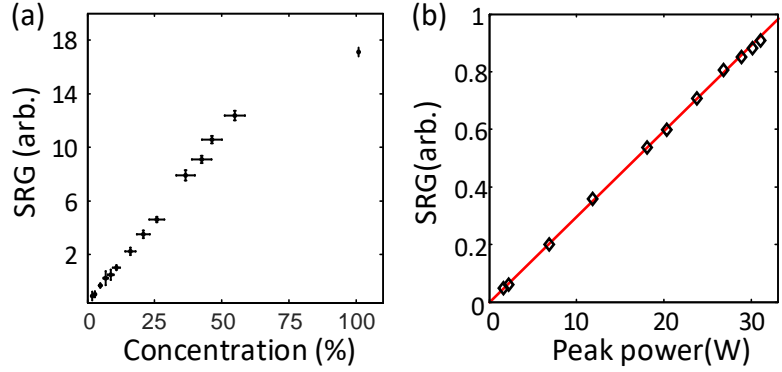

Fig. 3. Experiment results of dynamic range and linearity. (a) SRG signal for different concentrations of hydrogen balanced with nitrogen. The pump pulse width is 18-ns, the peak power delivered to the HC-PCF is 30 W. SRG signal was obtained by use of an oscilloscope with 200 averages. The error bars of concentration are enlarged for 3 times for clarity reason. (b) SRG of pure hydrogen as a function of peak power level of 18-ns pump pulse.

*3.3 Distributed sensing with high spatial resolution.*

To demonstrate distributed hydrogen detection with a higher spatial resolution, we used a 15-m-long HC-PCF sample as shown in Fig. 4(b) and 1-ns pump pulse. Pure hydrogen was pressurized into the HC-PCF through the air-gap at joint C with a pressure of ~0.7 bar above the atmosphere. The measured SRG distribution along the sensing HC-PCF as a function of time is shown in Fig. 4(a). SRG signal of hydrogen is clearly observable around $z = 7.5$ m and increases with gas loading time. The gain trace at 18 s is shown in Fig. 4(b) and the full width at half maximum (FWHM) of the gain peak around $z = 7.5$ m is ~75 cm. This value is larger than the spatial resolution calculated with $\Delta = \tau c/2 = 15$ cm and is believed to be mainly determined by the length of HC-PCF that is filled with hydrogen. Further increase in the pump power level would enhance the gain signal and allow measurement to be performed at an earlier stage (i.e., shorter length of HC-PCF that is filled with hydrogen) to demonstrate the pump pulse-width limited spatial resolution.

It should be stated that for the 18-ns pump pulse, the SRG can be determined by the gain factor in the steady-state regime. For short pump pulse (e.g., 1 ns), the coherence dephasing needs to be considered and SRG should be described by the backward Maxwell-Bloch equations [21]. For $S_0(0)$ transition of pure hydrogen, the dephasing time is ~3.3 ns at 1 atm. However, the spatial resolution of our distributed gas sensor is not limited by the dephasing time and could be further improved by using shorter pump pulses.

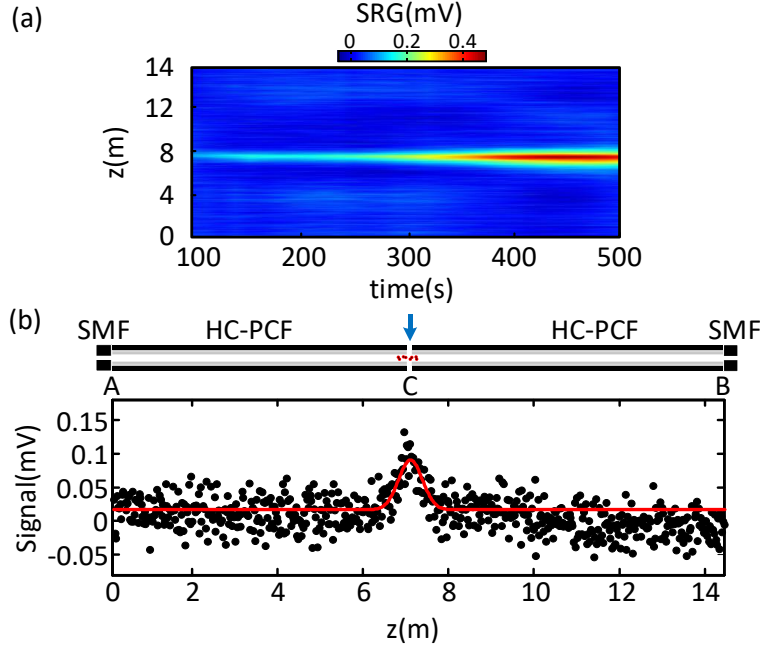

Fig. 4. Measured Raman gain trace along a 15-m-long HC-PCF with 1 ns pump pulse. (a) The Raman gain trace as a function of time during the hydrogen loading process. The peak power of pump is ~10 W in the HC-PCF. (b) Schema of the HC-PCF sample and the SRG trace at 18 s showing the hydrogen gain signal around z = 7.5 m. The local variance of the raw data is described by the error bars. Gaussian-fit of the gain profile around z = 7.5 m gives a FWHM of 75 cm. The blue arrow indicates the location of hydrogen loading. The construction of the HC-PCF sample is described in the Appendix C.

*3.4 Measurement of pressure distribution along the HC-PCF*

The distributed sensing system was also tested for the measurement of gas pressure distribution along the HC-PCF. A 13-m-long HC-PCF was used as the sensing sample (see Appendix D: Fig. 8(c) for details). The two ends of the HC-PCF sample (labeled as A and B) were applied with gas pressure of ~4 bar and 1 bar. The measurement was conducted after the pressure gradient in the HC-PCF reaches equilibrium. The wavelength of pump laser is tuned across the Raman gain spectrum of hydrogen while we keep the probe wavelength unchanged. The distributed SRG signal decreases as the wavelength of pump laser was tuned away from the Raman gain spectrum of hydrogen, as shown in Fig. 5(a). The inset in Fig. 5(b) shows the measured SRG spectrum around z =7 m. The gain spectrum of pure hydrogen can be fitted with a Lorentzian lineshape, from which the SRG linewidth can be determined. The backward SRG linewidth was used to decode the pressure distribution according to the linewidth we measured at different pressure (Appendix B: Fig. 6(a)). The results area shown as the dots in Fig. 5(b). Theoretically, the pressure distribution $P(z)$ along the HC-PCF with different gas pressure at the ends may be described by

$$P(z) = \sqrt{P_A^2 + \frac{z}{L}(P_B^2 - P_A^2)} \quad (1)$$

where $L$ is the length of HC-PCF, $P_{A,B}$ the gas pressure at the ends of HC-PCF (A and B), and $z$ the position respect to A. The experimental result fits well with the theoretical calculation, as shown in Fig. 5(b). However, due to interference of the SRG of SMF pigtails, the SRG spectrum of $S_0(0)$ transition is not accurately acquired near the ends of the HC-PCF sample.

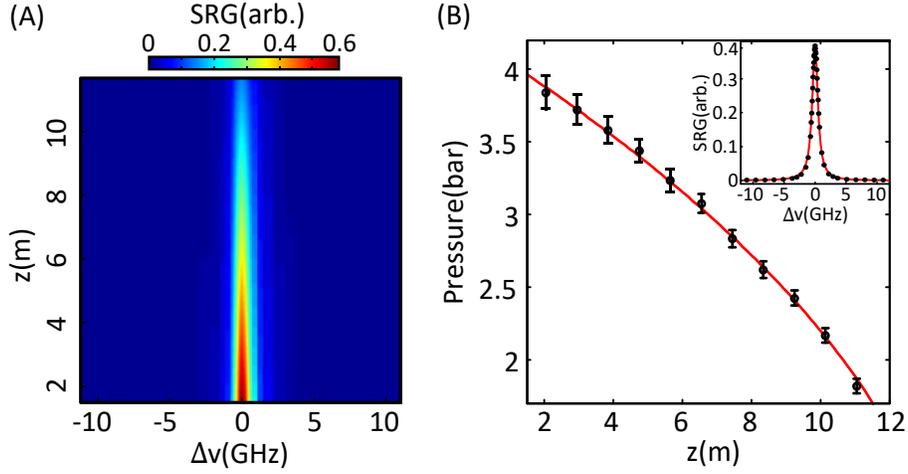

Fig. 5. Results of distributed gas pressure measurement. The peak power of pump pulse is 30 W with a pulse width of 18-ns. The SRG spectrum data were recorded by an oscilloscope with 200 averages. (a) The measured backward SRG spectrum along the HC-PCF. (b) The pressure distribution recovered from the SRG linewidth and the theoretical calculation based on Equation (1) with $P_A$ = 4.2 bar and $P_B$ = 1 bar. The inset shows a measured SRG spectrum (dot) around z = 7 m and Lorentzian fit (line).

## 4. Discussion

In summary, we have demonstrated all-fiber, label-free optical hydrogen sensors with point and distributed sensing capability. For the point sensor, a NEC of ~20 ppm was achieved. For the distributed hydrogen sensor, a NEC of 833 ppm with 2.7 m spatial resolution was demonstrated over a sensing length of 100 m at atmospheric pressure. The response time of the distributed gas sensor is less than 60 s with a dynamic range of more than three orders of magnitude. A higher spatial resolution of 75-cm was demonstrated over a sensing length of 15 m. Compared with previous distributed optical fiber hydrogen sensors (summarized in Table 1), the sensor reported here is the first label-free optical fiber distributed hydrogen sensor. It has high detection sensitivity and selectivity, large dynamic range, long sensing range and good spatial resolution.

Operating in the near infrared allows the use of cheaper and more compact photonic components compatible with standard telecom optical fibers, making it possible to develop cost-effective all-fiber, label-free distributed gas sensors. Within the transmission window of the HC-1550-06 fiber, the Raman transitions of several gases such as $H_2$, $O_2$ and $N_2$ are accessible, enabling multi-component gas detection. By using a hollow-core fiber (e.g., Kagome [24] or single-ring anti-resonant fiber [25, 26] with broader transmission band, vibrational or rotational Raman transitions of a wider range of gases may be accessed, and allowing the detection of many gases with Raman active transitions.

As shown in Fig. 3(a), the amplitude of SRG is proportional to the peak power of pump pulse and hence further enhancement of SRG signal is possible by simply increasing the peak power level. Numerical analysis using finite-element analysis shows that most of the light power of the fundamental mode in the commercial HC-1550-06 fiber is in the hollow core and cladding holes, and only a very small fraction (~0.4%) of the light is in the glass part. This results in a much higher threshold for nonlinear effects and optical damage, and allows the use of higher pump power to achieve higher detection sensitivity.

SRS is a very fast process that could enable a sampling rate of tens of megahertz. It is theoretically limited by the dephasing and depopulation times of hydrogen, which is on the order of nanoseconds at 1 atm. In our experiments, we demonstrated an effective sampling rate of 6 Hz over a sensing length of 100 m, limited by the number of averages used to improve the

SNR of SRG signal. With a higher power pump, a sufficiently high SNR could be achieved without the need for many averages. This would allow significantly higher sampling rate, which would enable fast dynamic fluidic analysis in the HC-PCF.

**Table 1. Comparison of Distributed Hydrogen Detection Systems**

The response and recovery time are transformed to a uniform standard as we defined above. [a]: [9] demonstrated a multi-point sensor with three 15-cm long sensors. [b]: The response and recovery time are roughly estimated from the data in Fig. 7 of [9]. [c]: The sensitivity is not stated but experiments show that 1% of hydrogen is detectable in [9] and [11]. [d]: theoretically estimated response and recovery time in [10] [e] : The response time of the gas sensor with heating (without heating) estimated from the data in Fig. 4 of [11].

| Technique | Label-free | Spark hazard | Sensing length | Spatial resolution | Response time | Recovery time | Sensitivity |
|---|---|---|---|---|---|---|---|
| **OTDR** [9] | No | No | $3 \times 15$ cm[a] | N. A. | A few hundred seconds[b] | | < 1% [c] |
| **LPG** [10] | No | No | 2.5 m | 10 cm | 1.3 min[d] | 4.3 min[d] | 1% |
| **OFDR** [11] | No | Yes | 2 m | 1 cm | 3 min (8 min)[e] | Not stated | < 1% [c] |
| **This work** | Yes | No | 100 m | 2.7 m | 46 s | 54 s | 833 ppm |

We demonstrated distributed hydrogen sensing over a 100-m-long HC-PCF with micro-channels drilled over a length of 2.2 m. It is feasible to fabricate a large number of micro-channels along the entire length of the HC-PCF for real time distributed gas detection. With a femtosecond laser, we have demonstrated the fabrication of hundreds of micro-channels with an average loss below 0.01 dB per channel [27]. Observation of the fabrication process showed that most of the micro-channels have negligible contribution to the fiber loss and a few poorly made micro-channels raised the level of the average loss, which is believed caused by the scrap generated during the femtosecond laser drilling process. By improving and automating the fabrication process, mass production of micro-channels with lower loss is possible, which would enable fast distributed gas sensing over the entire length of the fiber. With the newly developed single-ring anti-resonant hollow-core fibers, which have a simpler structure and better mode quality [25, 26], micro-channels or lateral cut along the entire length of the fiber could be made with negligible loss added to the fibers [28]. The SRG based distributed sensing principle could also be implemented with sub-wavelength exposed-core fibers [29, 30] in which the evanescent fields of the pump and probe interact with the surrounding gas molecules, avoiding the problems of fabricating micro-channels. The suspended-core fibers have a wider transmission window and would allow a wider range of gases to be detected.

## Appendix A: preparation of hydrogen sample with different concentrations.

The different concentrations of hydrogen were prepared by mixing pure hydrogen and nitrogen with different flow rate at atmospheric pressure. The gas mixture is filled in a gas chamber with a size of 20 cm × 30 cm × 1 cm. To prepare hydrogen with concentration of 1%, 2%, 4%, 6%, 7.9%, 10%, 15.2%, 20%, the flow rate of pure hydrogen was set to 5 sccm (standard cubic centimeter per minute) and the flow rate of pure nitrogen was set to 495, 245, 120, 78, 58, 45, 28, 20 sccm. The time of gas filling into the gas chamber was 4, 11, 15, 22, 29, 36, 55, 72 minutes, respectively. After the gas filling process, the gas chamber was sealed.

## Appendix B: Raman gain factor of hydrogen in HC-PCF

The Raman gain factor of hydrogen is polarization dependent. It takes the maximum value when the pump and probe beams have opposite circular polarizations and is given by [31]

$$g = \frac{8}{5} \frac{\pi^2 \omega_S}{c^2 n_S^2} \frac{(J+1)(J+2)}{(2J+1)(2J+3)} \frac{\gamma^2}{\hbar \Gamma} \Delta N \qquad (2)$$

where $\Gamma$ is the Raman linewidth, $c$ the speed of light in vacuum, $h$ the Planck's constant. $n_S$ is the refractive index at the Stokes frequency $\omega_S$, and is approximately 1 in HC-PCF. $\gamma$ is the off-diagonal element of the anisotropic polarizability [32], and $\Delta N$ is the number density difference of hydrogen molecules between J = 0 and J = 2. The linewidth of the backward Raman scattering is much larger than that of the forward Raman scattering due to Doppler broadening [33]. The forward linewidth (half-width-half-maximum, HWHM) of $S_0(0)$ transition of hydrogen is 48 MHz, calculated from the diffusion model (with parameters from [34]).

We measured the backward Raman linewidth of hydrogen with a 15-m-long HC-PCF filled with different concentrations of hydrogen balanced with nitrogen. The HC-PCF sample was made by butt coupling the HC-1550-06 fiber to SMF pigtails at both ends, as shown in Fig. 8(b). The butt coupling joints A and B were sealed respectively in T-shaped tubes being connected to pressurizing devices. The gas sample was pressurized into the HC-PCF through the T-shaped tube. The experiment setup is similar to that shown in Fig. 7, but a polarization controller instead of a polarization scrambler was used for the pump beam. The duration of pump pulse was 18 ns. Figures 6(a) and (b) show respectively the backward Raman linewidth and line shift measured with the HC-1550-06 fiber sample filled with gases with different hydrogen concentrations and pressure levels. The linewidth shows collisional broadening and narrowing (also known as Dicke narrowing) [33] as expected. The backward Raman linewidth of $S_0(0)$ transition of pure hydrogen is 1080 MHz at 1 atm. By use of Eq. (2), the forward and backward Raman gain factors were calculated to be 0.1 cm/GW and 4.5 cm/TW, respectively. If we consider the polarization scrambled pump as a natural incident pump light [35], the Raman gain factor would be 0.059 cm/GW and 2.6 cm/TW, respectively. Besides, as the number density difference is proportional to the hydrogen concentration $C$ by $\Delta N = \kappa C N_{tot}$, where $N_{tot}$ is the total number density and $\kappa \approx 0.1$ is a constant ratio calculated from the Boltzmann distribution at 296 K for $S_0(0)$ transition. Figure 6(a) shows that the backward Raman linewidths for 4% and 20% hydrogen are almost the same at 1 atm. Hence, the Raman gain factor of hydrogen balanced with nitrogen is expected to have a linear relationship with respect to the hydrogen concentration below 20%. This agrees well with the experimental test of the linearity, as shown in Fig. 3(a) in the main paper.

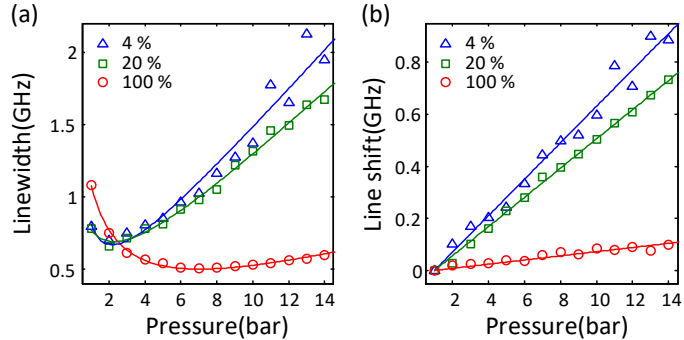

Fig. 6. Measured backward Raman linewidth and line-shift at 296 K. (a) SRS linewidth as functions of gas pressure for three different hydrogen concentrations. (b) Measured pressure induced Raman shift for different hydrogen concentration.

## Appendix C: setup for distributed hydrogen sensing based on backward SRS

The experiment setup for distributed hydrogen sensing is shown in Fig. 7. A distributed feedback semiconductor laser was used as the pump laser and an external-cavity diode laser was used as the probe laser. We use a polarization scrambler (PS) to eliminate the polarization fading effect since SRS is polarization dependent. The pump pulse is generated by using optical intensity modulator (IM) and is further amplified by an Erbium-doped fiber amplifier (EDFA).

Different IMs were used in different experiments. We use an acoustic-optic modulator (AOM) as the IM to generate the 18-ns pump pulses and an electro-optic intensity modulator (EOM) as the IM to generate 1-ns and 2-ns pump pulses. An optical filter centered at the pump wavelength filters out the amplified spontaneous emission (ASE) of the EDFA. A pump filter (PF) is used before photo-detection to filter out the residual pump light, and the time trace of the SRG signal is recorded and averaged with an oscilloscope.

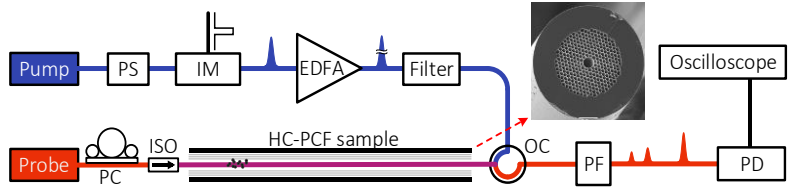

Fig. 7. Experimental setup for distributed hydrogen sensing. The inset picture shows the cross section of HC-1550-06 fiber. PS, polarization scrambler; IM, optical intensity modulator; EDFA, erbium-doped fiber amplifier; OC, optical circulator; ISO, isolator; PC, polarization controller; PF, pump filter; PD, photodetector.

**Appendix D: HC-PCF sample preparation**

Two types of HC-PCF samples were used in our gas detection experiments. One sample was fabricated by fusion splicing a 100-m-long sensing HC-1550-06 fiber to standard SMF pigtails at both ends. This HC-PCF sample is shown in Fig. 1(a) in the main paper and was used in the distributed sensing system to test the lower detection limit, response time and dynamic range. For the purpose of gas ingress/egress, 56 pairs of micro-channels (i.e., 112 micro-channels) were drilled over a 2.2-m region around the location 88 m along the 100-m-long sensing HC-PCF. The drilling was done with an 800-nm femtosecond laser micro-machining system [27]. The separation between the adjacent pairs of micro-channels is ~4 cm, and the spacing between the two channels forming a pair is 200 µm. The total loss of the HC-PCF sample measured from the input to output SMF pigtail is ~7 dB and ~8.7 dB before and after the drilling of the micro-channels. The average loss of a single micro-channel is estimated to be ~0.015 dB over the wavelength from 1530 to 1620 nm.

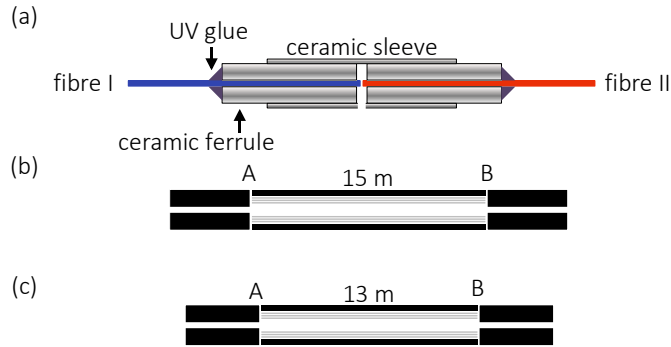

Fig. 8. HC-PCF samples. The HC-PCF is HC-1550-06 fiber. SMF is Corning SMF-28e fiber. (a) Method of butt coupling two fibers. The two fibers (fiber 1 and fiber 2) could be Corning SMF-28e fiber or HC-1550-06 fiber. (b) The 13-m-long HC-PCF sample used for distributed pressure sensing.

Three other HC-PCF samples were fabricated by butt coupling HC-PCF to SMFs and between HC-PCFs. These HC-PCF samples were used to test the performance of gas ingress/egress (the sample shown in Fig. 8(b)), for distributed hydrogen sensing with higher spatial resolution (Fig. 4(b)) and distributed pressure measurement (Fig. 8(c)). The method of butt coupling is shown in Fig. 8(a). Fibers are inserted into the ceramic ferrule and then assembled with a ceramic sleeve. A small gap of ~5 µm between the butt coupled fibers to allow gas flow into/out of the sensing HC-PCF. Fiber I and fiber II shown in Fig. 8(a) can be HC-PCF or SMF. Butt coupling joints were sealed within T-shape tubes that were connected with pressurizing devices for gas ingress/egress. A butt coupling joint introduces a typical loss of 2-3 dB in the sample we made.

## Appendix E: monitor the process of gas ingress/egress

The gas filling and purging process can also be monitored using our distributed hydrogen sensing system. The experiment setup is similar to that shown in Fig. 7, but we use a polarization controller instead of a polarization scrambler. First, we monitored the gas filling process of the 15-m HC-PCF sample (Fig. 8(b)) with 18-ns pump pulses. The 15-m HC-PCF sample was initially filled with air. We then load pure hydrogen with a pressure of 2 bar at point A, and point B was exposed to atmosphere. As shown in Fig. 9(a), the hydrogen gradually fills the whole HC-PCF sample. The SRG signal shows a gradient due to the pressure gradient and hydrogen concentration gradient along the HC-PCF. After the filling process reached equilibrium, both ends (i.e., points A and B) of the HC-PCF were exposed to atmosphere at around t ≈ 15,000 s. As shown in Fig. 9(a), the SRG signal reduces quickly and becomes almost the same along the entire HC-PCF within a short period of time.

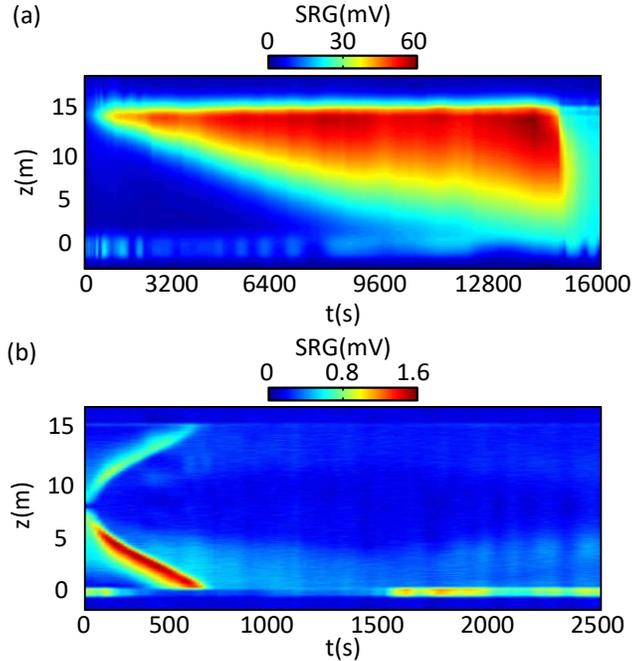

Fig. 9. Monitoring the gas filling and purging process. (a) The measured hydrogen filling process with an 18-ns pump pulse and the 15-m-long HC-PCD sample shown in Fig. 8(b). (b) The measured hydrogen purging process with a 2-ns pump pulse and the HC-PCF sample in Fig. 4(b).

Second, we monitored the gas purging process using the HC-PCF sample shown in Fig. 4(b), with a pump pulse width of 2-ns. The HC-PCF sample was initially filled with pure hydrogen. We then loaded pure nitrogen with a pressure of 4 bar at point B, and points A and C were exposed to atmosphere. As demonstrated in Fig. 9(b), the nitrogen 'pushes' out most of hydrogen in the HC-PCF in ~600 s. The SRG signal of hydrogen initially increased after we applied nitrogen with a pressure of 4 bar. The main reason for this is that the higher-pressure nitrogen compresses the hydrogen in the HC-PCF from 1 atm to a higher pressure. The SRG peak moved toward the two ends (i.e., points A and C) and eventually almost disappeared after 600 s, showing most of the hydrogen inside the HC-PCF has been pumped out by the higher pressure nitrogen gas.

## Funding

Research reported in this article is supported by Hong Kong SAR government through GRF grant PolyU152603/16E and through NSFC grant 61535004.

## Acknowledgments

The authors thank Chao Lu, Chao Jin, Nan Guo from the Communications Research Group at the Hong Kong Polytechnic University and Yiping Wang, Zhe Zhang from the Guangdong and Hong Kong Joint Research Centre for Optical Fiber Sensors at Shenzhen University for providing some of the components and equipment used for the experiments. The authors also thank Haihong Bao and Yuechuan Lin in our group for useful discussion.